\documentclass[conference]{IEEEtran}
\IEEEoverridecommandlockouts
\usepackage{cite}
\usepackage{amsmath,amssymb,amsfonts}
\usepackage{algorithmic}
\usepackage{graphicx}
\usepackage{textcomp}
\usepackage{xcolor}
\usepackage{stfloats}
\usepackage{url}
\usepackage[utf8]{inputenc}
\usepackage{fourier} 
\usepackage{array}
\usepackage{makecell}
\usepackage{multirow}
\usepackage{tabularx}
\mathchardef\mhyphen="2D % Define a "math hyphen"

\newcommand\norm[1]{\left\lVert#1\right\rVert}
\graphicspath{{figures/}}

\def\BibTeX{{\rm B\kern-.05em{\sc i\kern-.025em b}\kern-.08em
    T\kern-.1667em\lower.7ex\hbox{E}\kern-.125emX}}
\begin{document}

\title{Color Universal Design Neural Network for the Color Vision Deficiencies}

\author{\IEEEauthorblockN{Sunyong Seo}
\IEEEauthorblockA{\textit{Department of Media, Soongsil University} \\
Seoul, Republic of Korea \\
tjxhwl@gmail.com}
\and
\IEEEauthorblockN{Jinho Park}
\IEEEauthorblockA{\textit{Global School of Media, Soongsil University} \\
Seoul, Republic of Korea \\
c2alpha@ssu.ac.kr}
}

% \author{\IEEEauthorblockN{Anonymous}
% \IEEEauthorblockA{\textit{Placeholder} \\
% Placeholder \\
% Placeholder}
% \and
% \IEEEauthorblockN{Anonymous}
% \IEEEauthorblockA{\textit{Placeholder} \\
% Placeholder \\
% Placeholder}
% }

\maketitle

\begin{abstract}
Information regarding images should be visually understood by anyone, including those with color deficiency. However, such information is not recognizable if the color that seems to be distorted to the color deficiencies meets an adjacent object. The aim of this paper is to propose a color universal design network, called CUD-Net, that generates images that are visually understandable by individuals with color deficiency. CUD-Net is a convolutional deep neural network that can preserve color and distinguish colors for input images by regressing the node point of a piecewise linear function and using a specific filter for each image. To generate CUD images for color deficiencies, we follow a four-step process. First, we refine the CUD dataset based on specific criteria by color experts. Second, we expand the input image information through pre-processing that is specialized for color deficiency vision. Third, we employ a multi-modality fusion architecture to combine features and process the expanded images. Finally, we propose a conjugate loss function based on the composition of the predicted image through the model to address one-to-many problems that arise from the dataset. Our approach is able to produce high-quality CUD images that maintain color and contrast stability. The code for CUD-Net is available on the GitHub repository\footnote{ \url{https://github.com/ZombaSY/CUD-NET-release}}.

\end{abstract}

\section{Introduction}

Color blindness is a visual impairment that affects a significant percentage of the population. Specifically, the red-green type of color blindness is prevalent among individuals of Northern European descent\cite{Won11}, affecting approximately 8\% of males and 0.5\% of females. This form of color blindness is the most common, followed by blue-yellow color blindness and total color blindness. To address this issue, we have utilized deep learning model to develop color universal design (CUD) images that are compatible with individuals who have red color deficiencies (protanopia) and green color deficiencies (deuteranopia). It is important to note that the degree of individual color deficiency varies and, as such, the severity of protanopia and deuteranopia may differ. Our study focuses on developing CUD images that cater to the specific needs of individuals with the protanopia and deuteranopia vision.

\begin{figure}[htp]
  \centering
  \includegraphics[width=.9\linewidth]{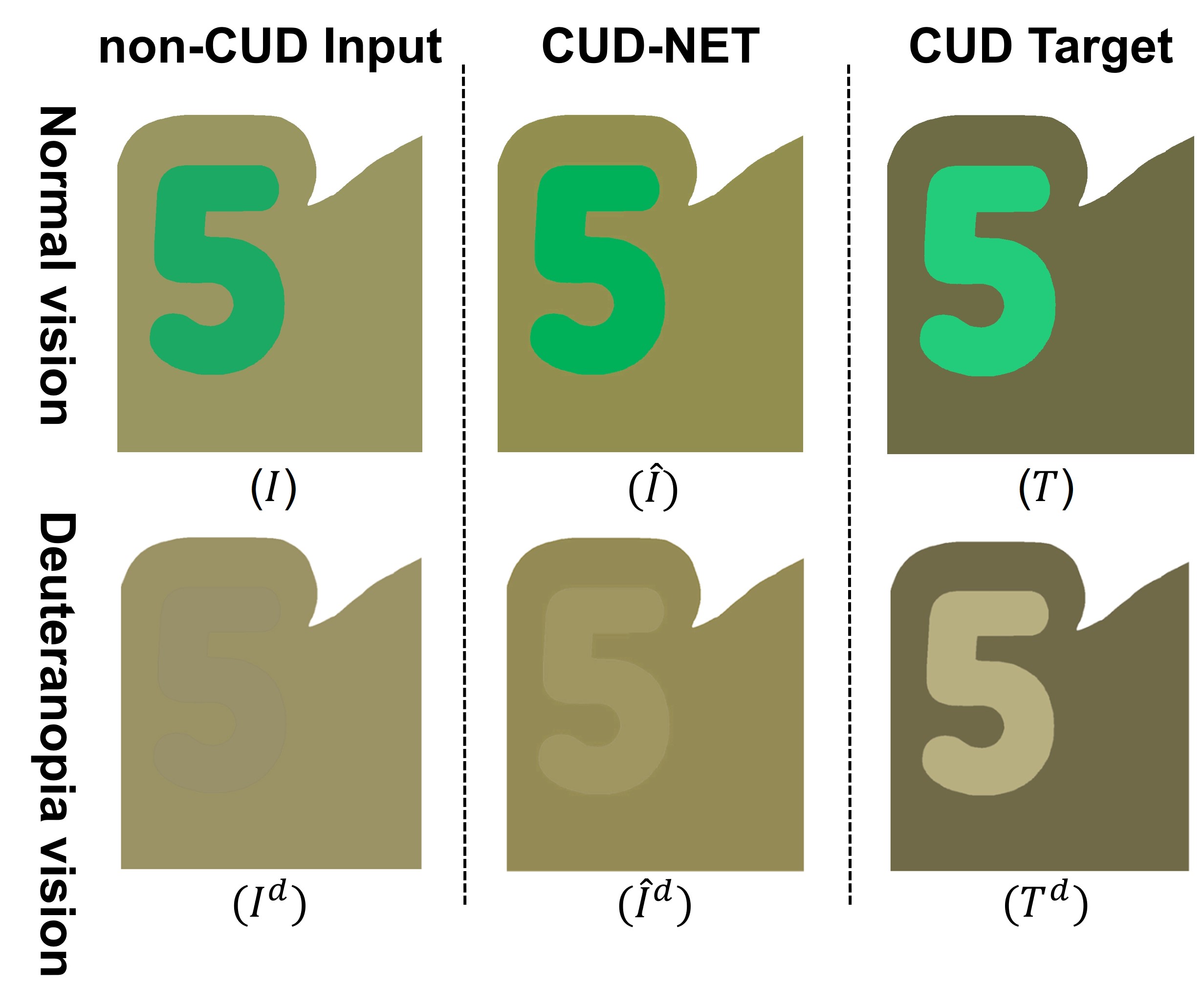}
  \caption{
  Comparisons of the non-CUD image, CUD-Net's predicted image, and CUD image. The top row is represented in normal vision, and the bottom row in deuteranopia vision. Numeral '5' appears on our predicted image in deuteranopia vision and complies with color preservation.}
  \label{fig:02}
\end{figure}

Previous research has investigated color vision deficiencies using wearable devices \cite{VZCR20}. However, such studies incur significant time and cost overheads on hardware devices. To address this issue, we propose a novel approach that utilizes deep learning-based image enhancement methods to generate color universal design (CUD) images in real-time. Our method enhances the visibility of colors for individuals with specific color deficiencies. For example, consider the image $I$ shown in the top-left corner of Figure \ref{fig:02}. Individuals without color deficiencies can easily recognize the numeral ‘5’ in this image. However, those with deuteranopia vision may struggle to differentiate the numeral from the surrounding colors, making it difficult to distinguish the boundary of adjacent objects. We define non-CUD objects as those that are invisible to individuals with color deficiencies, and CUD objects as those that are distinguishable by both individuals with normal vision and those with color deficiencies. Our aim is to generate a CUD image $\hat{I}$ by applying a specific filter to the original image $I$. This filter highlights the CUD objects while ensuring that the surrounding objects are also visible to individuals with color deficiencies.

\begin{figure*}[tbp]
  \centering
  \mbox{}
  \includegraphics[width=.95\linewidth]{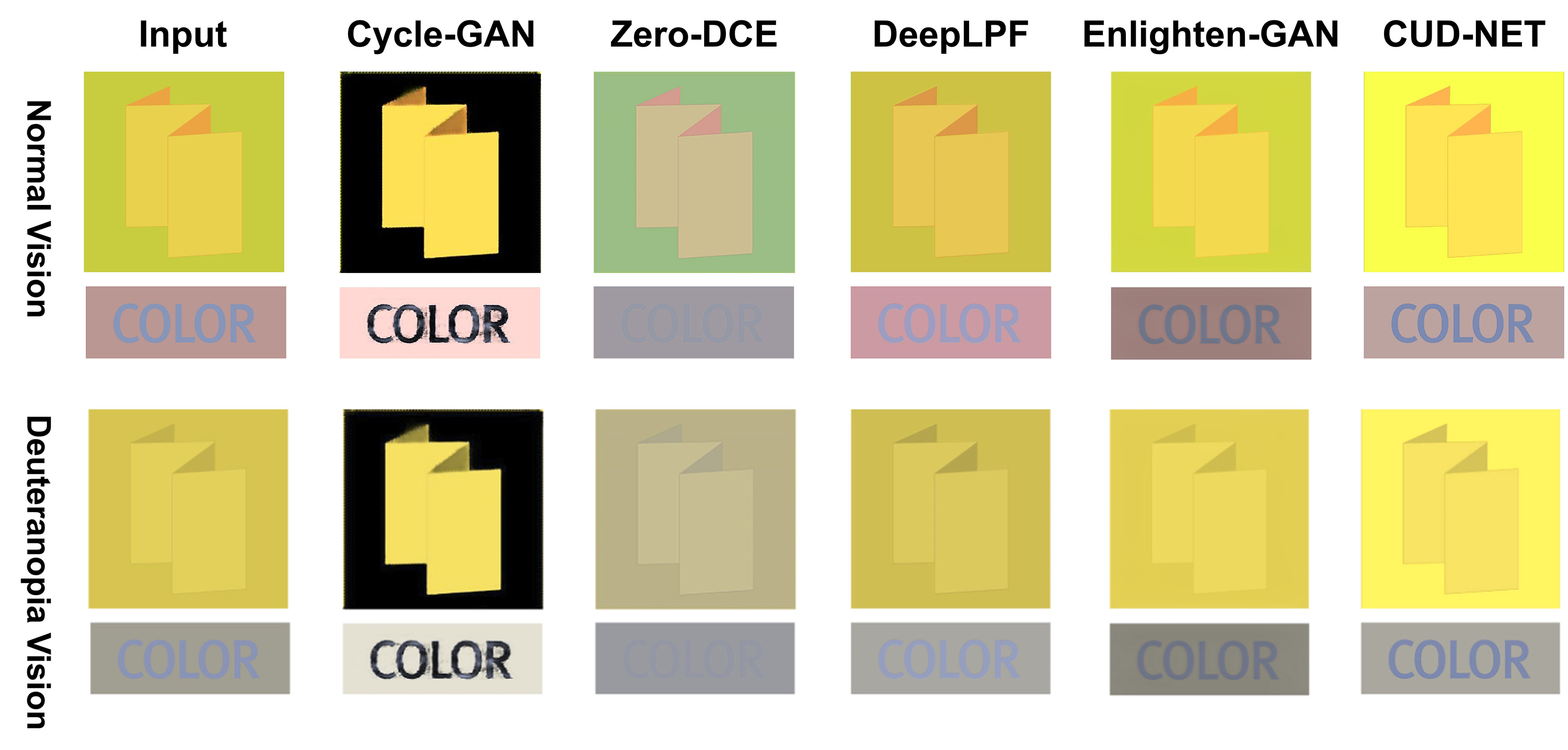}
  \caption{\label{fig:01}
           Predicted image conversion for comparison experiments in the training set. The items are arranged in descending order of evaluation, starting from the left. It's noteworthy that all experiments were conducted using the identical dataset, which had been refined by color experts based on specific criteria.}
\end{figure*}

Our objective is to apply the weakest possible filter to the CUD objects while preserving color fidelity. This requires a comprehensive understanding of objects in the image. Various studies\cite{ZhuMao2021} have been conducted on defining specific objects or areas in an image, ranging from classic principal component analysis (PCA) \cite{WEG87} to machine learning-based segmentation methods \cite{TSC20, ZGL*20, kirillov2023segment}. However, deep learning algorithms currently lack a complete understanding of all objects in the real world without additional contexts, as evidenced by the limited success of visual question answering for arbitrary questions about object interactions \cite{AHB*18, KZG*17, LYL*20}. As a result, we propose a CUD-network (CUD-Net) that expands the input image's feature space around color deficiency vision information and defines a robust neural filter. Our approach ensures that the resulting image is suitable for CUD while maintaining the source image's color preservation.

In this paper, we propose a CUD-Net to satisfy both color preservation and contrast of non-CUD objects (CUD suitability). We present four core contributions of CUD-Net.

\begin{itemize} 
\item \textbf{Dataset refinement criteria for CUD images} 
Our approach involves refining the training data into two distinct groups. The first group consists of simple color tone images that are based on the H and V values in the HSV color space. The second group comprises images with two or more non-CUD objects that are required to be distinguishable in publications.

\item \textbf{Image pre-processing for CUD-Net}
In order to enhance the input image, we carry out pre-processing to amplify the available information. More specifically, we reconstruct the input image, $I$, utilizing three distinct features that have been expanded, while simultaneously eliminating any extraneous noise.

\item \textbf{Multi-modality fusion architecture}
We define feature, fusion, and regression layers to handle pre-processed images. The three features from the feature-extracting layer are combined into one fusion feature. Finally, a filter is constructed by regressing the node point of the piecewise linear function or indicator of the filter.

\item \textbf{Conjugate loss function}
We proposes a loss function that has conjugate loss and is based on the predicted image generated by our model. This method is crucial as our data is faced with a one-to-many problem, wherein a specific color in the input image $I$ is mapped to multiple colors in the target image \(T\).

\end{itemize}

\section{Related Works}

\subsection{Algorithm for color vision deficiency}
As per the color deficiency survey\cite{RG19a}, various approaches have been employed to address color deficiency by filtering or scaling different color spaces such as LMS, RGB, HSX, CIE Lab, and YCC in sequence. Iaccarino et al.'s method\cite{IMPS06} involves scaling the pixel value that satisfies the threshold of R and G colors and rotating the hue value of the reference color. In contrast, SPRWeb\cite{FRGG13} aims to satisfy both the perceptual experience of color vision deficiency and the subjective experience, while preserving color and contrast. Their method uses a two-pass hill-climbing algorithm to optimize the objective function with four components: color naturalness, perceptual color contrast, subjective color naturalness, and subjective color contrast. While it has the advantage of being applicable to natural images, most of the existing recoloring algorithms fail to account for image complexity, leading to the same adjustment filtering being applied to considerably simple images (with only two color tones, as illustrated in Figure \ref{fig:02}).

\subsection{Image-to-Image translation based on generative models}
Generative adversarial networks (GANs) have become a popular tool for image translation tasks, including image generation, style transfer, and colorization\cite{KWK21, IZZE17}. Among various GAN models, Cycle-GAN\cite{PEZZ20} has shown superior performance in enhancing the contrast of non-color-under-deficient (CUD) objects, as depicted in Figure \ref{fig:01}. However, to achieve color preservation, it is crucial to prevent the input image from losing its original color, even in the worst case scenario where the object is completely black. To address this issue, Enlighten-GAN\cite{JGL*21} has been proposed, which improves stability and generates more reliable results in terms of color preservation. 

\subsection{Image enhancement based on neural filter estimation}
In contrast to GANs, some studies have explored the approach of scaling pixel values of images through neural filter estimation techniques\cite{WZF*19, DLT18, BCPS19}. Zero-DCE\cite{GLG*20} and DeepLPF\cite{MMM*20} are two such studies that aim to enhance low-light images by estimating pixel-wise and high-order filters for dynamic range adjustment using lightweight deep networks. While Zero-DCE focuses on providing a brighter visual display of input images, DeepLPF attempts to enhance image contrast while maintaining color preservation through the use of graduated elliptical and polynomial filters that are easy to understand for viewers. However, in our problem, applying contrast factor yields similar results to the input image in both visions, indicating an overly stable filter. This may be attributed to the inability to comprehend the interaction of objects, leading to the development of an overly stable filter. We conducted additional assessments of the transformer-based neural architecture, yet its outcomes exhibit poor color harmony, consistent with the observations made in \cite{PEZZ20}.As a consequence, we have developed our backbone network utilizing a neural architecture based convolutional neural network.

\section{Methodology}

\subsection{Overview}
\begin{figure}[htb]
  \centering
  \includegraphics[width=.9\linewidth]{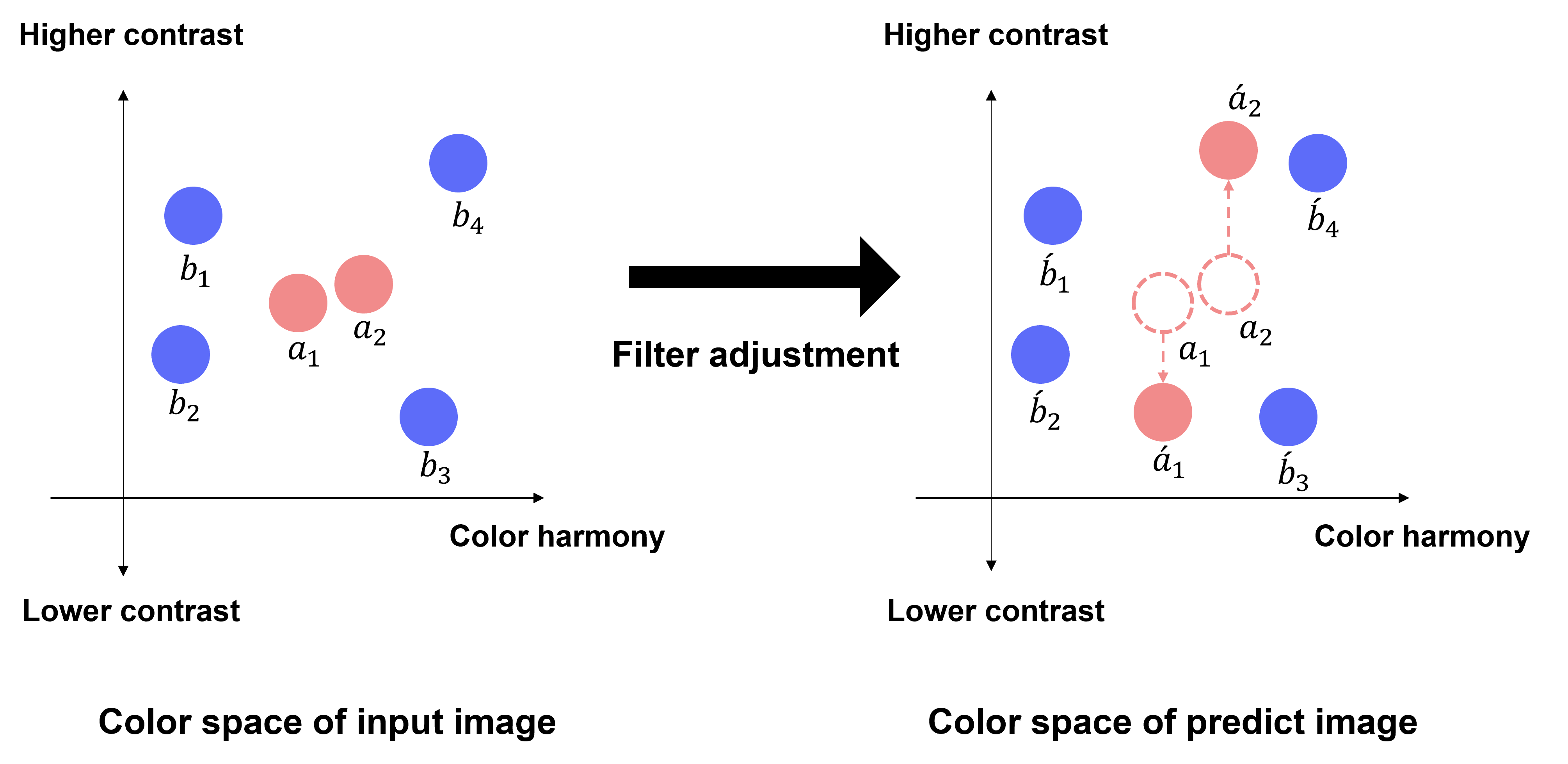}
  \caption{\label{fig:03}
           Ideal color conversion of the predicted image by balancing contrast and color preservation. The non-CUD object \(a\) should widen the gap compared to the input image and preserve its original color harmony, whereas the CUD object after filter adjustment \(\acute{b}\) preserves both contrast and color harmony.}
\end{figure}

In this study, we aim to generate an ideal predicted image that enhances the contrast of non-CUD objects while preserving the color of CUD objects in the input image. Specifically, we seek to map non-CUD object \(a\) to \(\acute{a}\) and maintain the color and contrast of CUD object \(b\), as exemplified in Figure \ref{fig:03}. However, since our neural filter operates on all pixels of the image, we are constrained to apply the same filter to both objects \(a\) and \(b\). Achieving a perfect match between the contrast and color of object \(b\) before and after the filter adjustment while maximizing the contrast of object \(a\) is challenging. To address this issue, we propose a deep learning-based regression technique that designs a specific filter for each image to maximize the contrast of object \(a\) while minimizing the impact on features of object \(b\).

Our proposed solution aims to maximize the contrast of the L channel values in the CIELab color space\cite{RG19b}. We have empirically verified that color deficiencies such as protanopia and deuteranopia, which are the most common types of color blindness, can differentiate between non-CUD objects when their L channel values are adjacent to each other. To illustrate this, in Figure \ref{fig:02}, the L channel value of the numeral '5' in image \(I\) is 61, and the surrounding color is also 61. This distinction is easy to perceive with normal vision but is ambiguous for individuals with deuteranopia when observing image \(I^d\). In contrast, for the CUD target images, \(T\) and \(T^d\), the L channel values have a difference of 75 for the numeral '5' and 45 for the surroundings, making it easy to distinguish between normal and deuteranopia vision. We refine our data pairs by defining a criterion that separates two invisible non-CUD objects using their L channel values, taking into account the characteristics of the data.

\begin{figure*}[tbp]
  \centering
  \mbox{}
  \includegraphics[width=.95\linewidth]{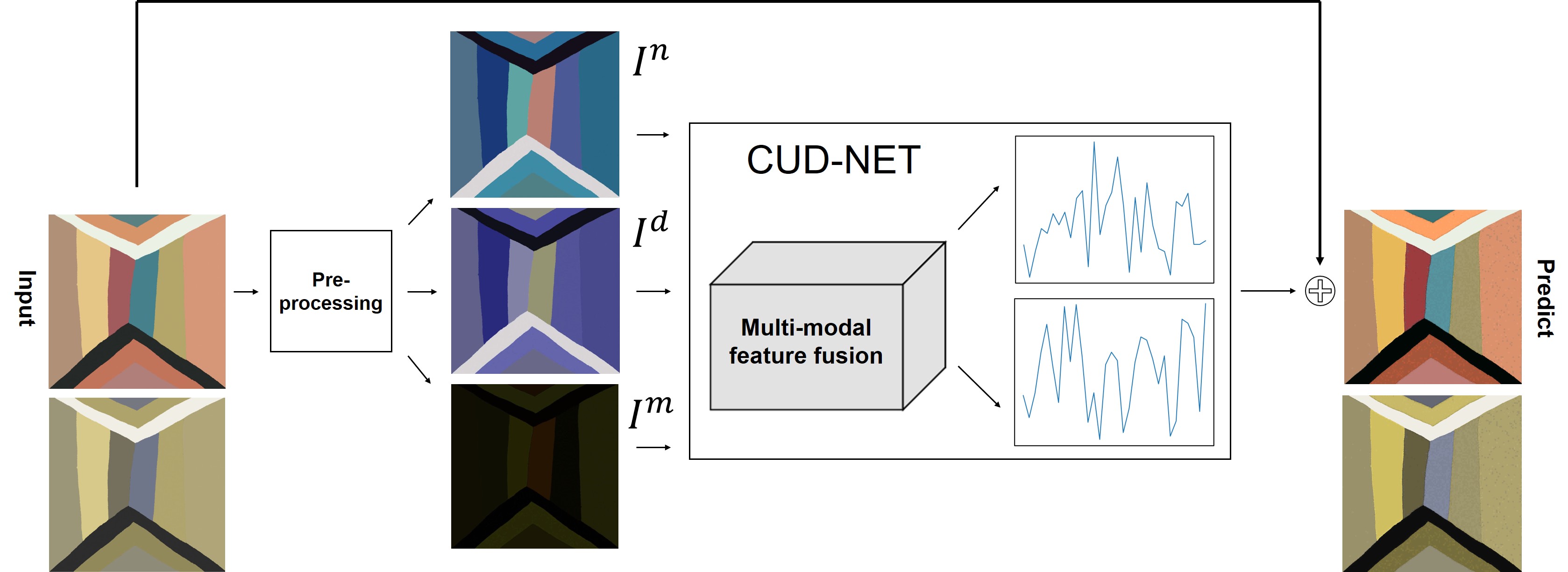}
  \caption{\label{fig:04}
           Overview of CUD image generation. The input is initially divided into three feature images prior to its passage through the CUD-Net. The CUD-Net, as depicted by the blue line in Figure \ref{fig:04}, proceeds to regress the node points of a piecewise linear function within the S and V channels. Ultimately, this filter is applied to the original input. On the right side of the predicted image, the content is distinguishable to individuals with both normal and deuteranopia vision. However, in the input image with deuteranopia vision, this discrimination is not feasible.}
\end{figure*}

In our proposed method, we address the issue of color preservation while maximizing the contrast between non-CUD objects through a conjugate loss function and a multi-modality fusion network. Although enhancing the contrast of L channel values between non-CUD objects is a crucial aspect, it is equally important to maintain the color fidelity of the input image. If we prioritize contrast enhancement over color preservation, we may end up with a filter that polarizes the color of the non-CUD objects to black and white, resulting in a loss of information in CUD objects. Hence, we aim to strike a balance between contrast enhancement and color preservation through an appropriate trade-off. We achieve this by incorporating a conjugate loss function and a multi-modality fusion network.

\subsection{Dataset refinement criteria for CUD images}
In our methodology, the training data were sorted into two categories. The first group consisted of vectorized images with two colors that were divided based on value V and hue degree H in the HSV color space\cite{HMKO19}. The second group comprised images with two or more objects that needed to be differentiated while preserving the color of non-CUD objects. The training data was categorized into approximately 1,600 different color combinations based on the same V, and then simulated with deuteranopia vision by converting them to the adjacent color family to ensure color preservation. All color conversions were scaled only within S and V in the HSV color space to achieve a minimum of 15 differences in the L channel value within the chosen non-CUD objects. The colors were combined with 10 essential hues and tones, and a similar color was simulated by converting it to deuteranopia vision. The essence of refining the training data is to prioritize color preservation, which enables our models to adhere to the same approach during the learning process.

\subsection{Image pre-processing}
The CUD-Net involves regressing the node points of a piecewise linear function and applying the final filter as a multiplication operation on the input image. However, it has been observed that performing multiplication with values less than 1 often results in a decrease in color saturation. When an image is devoid of any color inversion, the white color value converges to 1, which means that if the multiplication value falls within the range of [0, 1], the white color shifts towards black. To overcome this problem, we propose to invert the pixel value of the input image, which effectively ignores the multiplication operation for white values of 0. Additionaly, we employ Daltonization\cite{DTAA09} to compute map image \(I^m\) from the original RGB input images by calculating the difference value between images viewed with normal and deuteranopia vision. Increasing the channel on the stem block tend to generate predicted images that ignore the source color, leading to polarized colors similar to those produced by Cycle-GANs

\begin{equation}\label{eq:loss_function}
\begin{aligned}
I\ =\ \delta\left(\left\{\ I^n,\ I^d,\ \left|invert(I^n)-\ invert(I^d)\right|\ \right\}\right)
\end{aligned}
\end{equation}
\[
   where\ \ \ \delta\left(x\right) \begin{cases} 
        0, \ \ \ \ \ \ \ \ \ \ \ \ \ \ \ \ \ \ x\ <\ 0 \\
        x, \ \ \ \ \ \ \ \ \ \ \ \ 0\ \le\ x\ \le\ 1 \\
        1, \ \ \ \ \ \ \ \ \ \ \ \ \ \ \ \ \ \ x\ >\ 1 \\
        
        \end{cases}
\]

After applying color inversion \(invert(.)\), from the original RGB input image, \(I^n\), we generate image \(I^d\) with the aspect of deuteranopia vision. From these two generated images, we can obtain the absolute difference value to compose the map image. In Equation 1, the final input, \(I\), concatenated with \(9\times H\times W\) dimensions passes through the model. The \(\delta(.)\) clips the output to the range, [0, 1].

To generate the map image, we first apply color inversion to the original RGB input image, \(I^n\), resulting in the deuteranopia vision image, \(I^d\). The absolute difference value between these two images is then calculated to obtain the map image. To input this map image into the model, we concatenate it with the original input image, resulting in a tensor of dimensions \(9\times H\times W\) where H and W are the height and width of the input image, respectively. This concatenated tensor is then passed through the model, and the output is clipped to the range [0, 1] using \(\delta(.)\).

\subsection{Model Architecture}

%%% revision-start
\begin{figure*}[tbp]
  \centering
  \mbox{}
  \includegraphics[width=.95\linewidth]{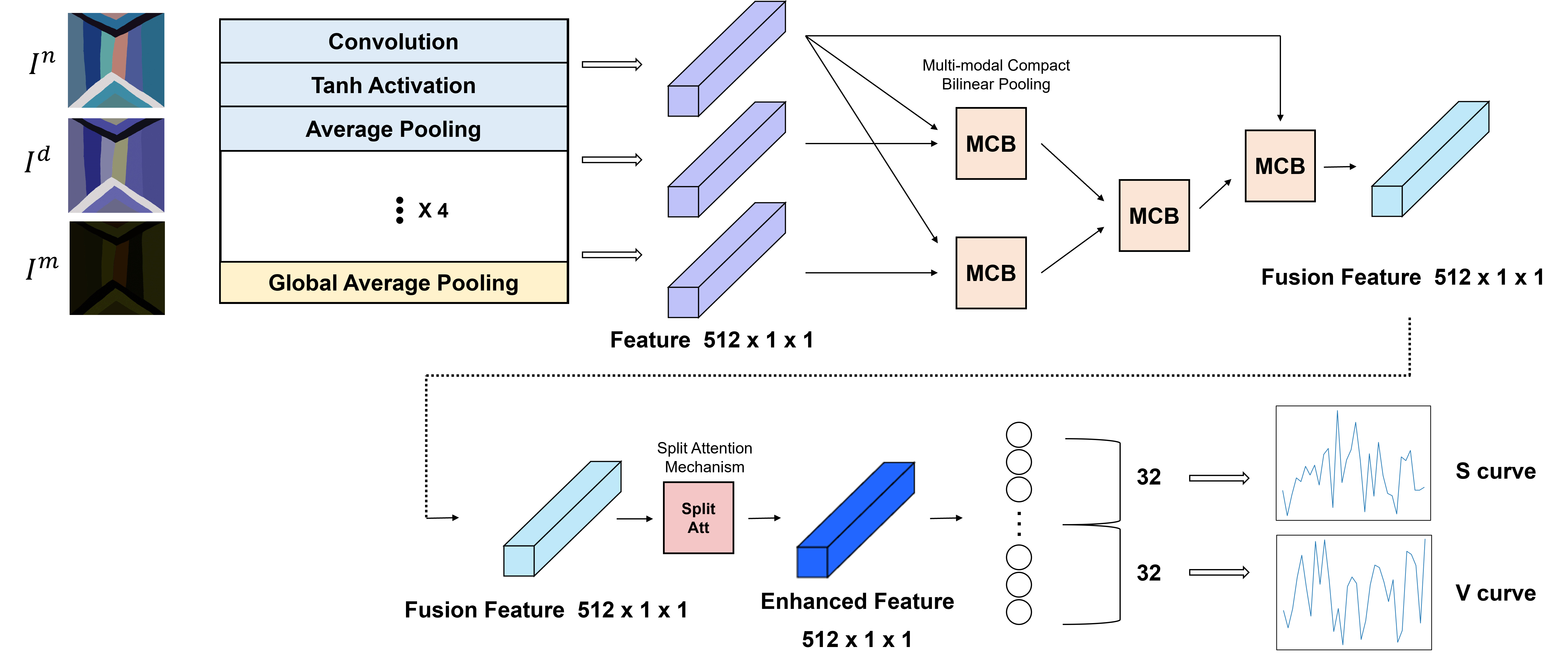}
  \caption{\label{fig:05}
        Structural overview of CUD-Net.}
\end{figure*}
%%% revision-end

As aforementioned, the CUD-Net model utilizes the node points of a piecewise linear filter function to predict the output image based on the input. Each node point value is used to perform a multiplication operation, resulting in the final predicted image. Figure \ref{fig:04} demonstrates how the input image \(I\) is processed by CUD-Net. The input is initially compressed into three feature blocks using convolution, pooling, and global pooling layers. The first three channels of the feature blocks serve as the primary input for the multiplication operation, while the remaining six channels act as additional features.

We employ a multi-modality architecture to extract expanded features. We convert three separate inputs into the HSV color space and use weight-sharing convolution layers with a kernel size of 3, a stride of 1, and a padding of 1 to extract a feature block corresponding to each input. To reduce dimensionality, we apply average pooling and use hyperbolic tangent as the activation function, as most output feature values fall within the range [-1, 1], allowing us to construct the node points. 
%%% revision-start
Each block is repeated four times to reach final features. 
%%% revision-end
To address the issue of unstructured image size in the inputs, we maintain the feature size of the last global pooling block instead of using average pooling, as proposed in previous work\cite{LCY14}.

%%% revision-start
As illustrated in Figure \ref{fig:05}, {\((I^n, I^d, I^m)\)} are encoded into each feature block using weight-sharing parameters. In our approach, we employ the multi-modal compact bilinear pooling gate (MCB)\cite{FPY*16} to combine these three stages multiplicatively. MCB aids in projecting feature representations into a higher-dimensional space through the use of Count Sketch\cite{CHARIKAR20043} and efficiently convolving both vectors by employing element-wise product operations in the Fast Fourier Transform space. To maintain the original image characteristics within the fusion layers, \(I^n\) is predominantly utilized within the MCB blocks, as it closely aligns with the essential properties of the original input \(I\). Given that MCB relies on stochastic selection, increasing the frequency of \(I^n\) would enhance its essential properties. The enhanced feature derived from the fusion process undergoes processing through the fully connected regression layer. In our experimental analysis, we ascertain that the optimal number for regressing the piecewise linear function is 64 points. These 64 points are evenly distributed into two halves, wherein the first half is employed to formulate the node points for channels S, and the latter half is for channel V within the HSV color space. These node points play a crucial role as scaling factors for generating the predicted image as outlined in Equation 2\cite{MMS19}.

%%% revision-end

\[S\left(I_i^{s,v}\right)=\ k_0+\ \sum_{m=0}^{M-1}{\left(k_{m+1}-\ k_m\right)\ \delta\left(\ MI_i^{s,v}-m\right)}\ \ \ \  (2)\]

The total number of node points, \(M\), and each pixel value of the S and V channels in the input image, \(I_i^{s,v}\), are multiplied by the slope of the actual regressed value, \(k_m\), the \(m\mhyphen th\) generated node point. The specific node points, \(M\), are scaled through a multiplication operation to the pixel value of the input image according to each node point.

\subsection{Loss function}
Our data presents one-to-many problems, where the relationship between input and target data is not unique. In other words, a color in the input image may correspond to multiple colors in the target image. To address this issue, we propose a loss function, denoted as \({L}\), that takes into account the potential and diversity of the predicted image. Specifically, \({L}\) is designed to capture the range of valid colors in the target image, as well as the distance between the predicted and target colors. We ensure differentiability of all equations included in the loss function. This approach allows for a more robust and flexible model that can handle the inherent ambiguity in the dataset. The dataset pair, \({(I_1, T_1),\ \ (I_2, T_2),\ \ldots,\ (I_n, T_n)}\), exemplifies the one-to-many dataset structures that motivate our loss function:

\[{L}=\ \sum_{i=1}^{N}{{Lab}_{loss}\left(V\left(\Phi\left({\hat{I}}_i\right)\right)\right)+\ H_{loss}\left({\hat{I}}_i\right)\ \ \ \ \  (3)}\]

\textbf{Stencil Masking.}
As outlined in the dataset refinement criteria, we do not perform color conversion on all areas of the target images. Only the areas that have color combinations that are invisible to deuteranopia vision, or non-CUD objects, undergo color conversion. As a result, the input image has regions with colors that are converted and not converted, which can also be referred to as CUD and non-CUD objects. To indicate the color boundary to the model, we introduced the stencil masking method.

\[\Phi\left({\hat{I}}_i\right)\ =\ {\hat{I}}_{ij}\ \ \mathrm{\cdot}\ I_{ij}\ \mathrm{\cdot}\ {\ T}_{ij}\ \ \ \ \ (4)\]

To identify the non-CUD objects in the input and target images, we generate stencil maps through logical AND operations on each pixel value, \(I_{ij}\) and \(T_{ij}\). These maps temporarily convert the pixel values into discrete integer values for the logical operation. By computing the stencil map, we can specify the non-CUD areas in the image. Using the predicted image, \({\hat{I}}_{ij}\), we can further refine the image using a logical OR operation in Equation 4. We apply a stencil mask to the CUD objects in the image, which allows us to exclude them from the neural filter computation. This approach is similar to how we refine the target image. Finally, the refined image is calculated using the loss function.

\textbf{CIELab Loss}
The CIELab channel loss function is employed to enhance color contrast for deuteranopia vision. In order to ensure stable brightness and contrast of the predicted image, we compute the MS\(\mathrm{\cdot{}}\)SSIM (multi-scale structural similarity\cite{WSB03}) of the L channel.

\[ {Lab}_{loss}\ =\  \norm{Lab\left(\hat{I}_i^{rgb}\right) - Lab\left(T_i^{rgb}\right) }_1\ + \ \ \ \ \ \ \ \ \ \ \ \ \ \ \ \ \]
\[ \ \ \ \ \ \ \ \ \ \ \ \ \ \ \ \ MS\mathrm{\cdot} SSIM\left(Lab\left({\hat{I}}_i^{rgb}\right),\ Lab\left(T_i^{rgb}\right)\right) \ \ (5) \]

The \(Lab\left(.\right)\) expression in Equation 5 returns the CIELab channel corresponding to the RGB channel. In computation, only L channels on Lab color space is used in calculation.

\textbf{Histogram Loss.}
To preserve the color of the input image in the output, we employ a histogram loss function that operates on the RGB channel. Unlike other loss functions that use only the L channel, we believe that preserving color in the RGB space produces superior results. By handling the RGB channel as a loss function, we achieve better color preservation than with the ab channels of the Lab color space.

\[H_{loss}=-\omega_{hist}\ \int N\left({\hat{I}}_i^{rgb};\ \sigma\right)-\ N\left(T_i^{rgb};\ \sigma\right) \ \ \ \ \ \ \  (6)\]

When using the L1 distance of the RGB channel pixel values as the loss function, we observed sensitivity to certain values and diverging gradients, making the experiment untrainable. To overcome this issue, we used the Gaussian expansion method proposed by Sandler et al.\cite{SAC*17}, represented by \(N(.)\), to derive a differentiable histogram loss function as described in Equation 6. The difference in the RGB channel of the differentiable histogram function is computed, and this can be modified to mean squared error or cosine similarity. The scaler, denoted as \(\omega_{hist}\), is inverse proportional with the size of the input image. This correlation arises due to the exponential growth of the histogram loss function as the number of pixels increases. By ensuring the similarity between the RGB channels of the predicted and target images, we can achieve color preservation.

\textbf{Conjugate Prediction.}
The target image is modified to include at least two colors when compared to the input image. However, as the model generates the predicted image using a neural filter, it is not predictable which color area will be modified. Hence, if the predicted image's color is shifted in the opposite direction or over-shifted concerning the target image, the loss will increase. Additionally, since the data pair does not correspond to a one-to-one mapping in a particular color, it is necessary to produce an alternative predicted image that adheres to the data pair's essence. We employ a conjugate prediction to estimate the loss function using the predicted image for potential color shifts.

\begin{figure}[htb]
  \centering
  \includegraphics[width=.9\linewidth]{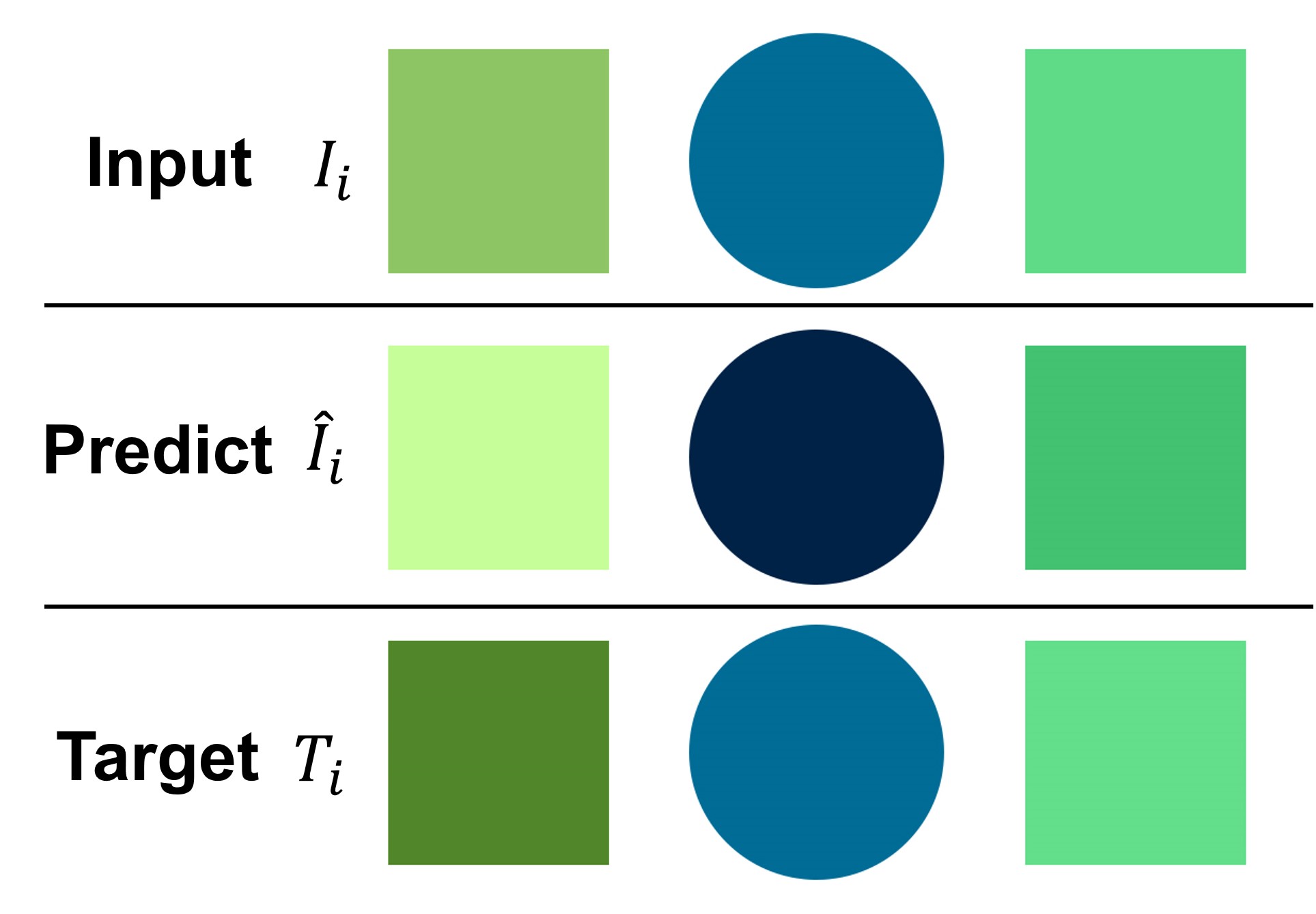}
  \caption{\label{fig:06}
           An example of conjugate prediction. In this scenario, the L value of the predicted image has been increased in the left rectangle and decreased in the right rectangle compared to the input image. Conversely, the target image has been decreased in the left rectangle and increased in the right rectangle compared to the input image. Despite satisfying the CUD condition within the predicted image, the loss function yields a notably high loss value. In response to such cases, we introduce a conjugate loss function.}
\end{figure}

The first case occurs when the predicted image has an excessive shift in color values compared to the target image. To address this, we refine the data pair by adjusting the L channel difference to a more reasonable value. This is illustrated in Figure \ref{fig:06}, in the order of left-most rectangle, circle, and right-most rectangle, \(I_i^L\) =\ \{74,\ 41,\ 79\}, \({\hat{I}}_i^L\ =\ \{97,\ 10,\ 70 \}\), \(T_i^L\ =\ \{50,\ 41,\ 80\}\) in L channel value, where the green rectangles on both ends have similar L channel values and belong to the same color family. We clip the excessive L channel value in the predicted image using Equation 7, which prevents any values from exceeding a certain threshold.
The second case is when the predicted image has an opposite shift in color values compared to the target image. In this scenario, using the complete neural filter results in a large mean squared error between \({\hat{I}}_i^L\) and \(T_i^L\). To overcome this issue, we generate an alternative predicted image using Equations 9 and 10. This allows us to adjust the predicted image while preserving its overall structure and features.

\[
   clip\left({\hat{I}}_{ij}\right) \ =\ \begin{cases} 
        max{\left({\hat{I}}_{ij},\ {T}_{ij}\right)},\ \ \ \ \ \ \ \ \ \ \ \ \ \ I_{ij}\ >{T}_{ij} \\
        min{\left({\hat{I}}_{ij},\ {T}_{ij}\right)},\ \ \ \ \ \ \ \ \ \ \ \ \ \ \ I_{ij}\ \le{T}_{ij}
        \end{cases}
        \ \ (7)
\]

\[{R}_1=\ 2I_{ij}-\ {\hat{I}}_{ij}, {R}_2=\ {\hat{I}}_{ij}\ \ \ \ \ \ (8)\]

\[V\left({\hat{I}}_{ij}\right) \ = \ argmin\left( \norm{clip \left(R_{1,2} \right) - {T}_{ij}}_2 \right) \ \ \ (9)\]

As described above, the thresholds are defined by the maximum and minimum values of each corresponding pixel position of \({\hat{I}}_i^L\) and \(T_i^L\). By computing the difference between the residual map and the input image, we induce two alternative images, \(R_1\) and \(R_2\). Subsequently, \({\hat{I}}_i^L\), with a smaller L2 distance is selected as an alternative predicted image in Equation 9, and it is finally computed with a loss function compared to the \(T_i^L\). The above equation establishes \(V\left(\Phi\left({\hat{I}}_i\right)\right)\) = \{54,\ 41,\ 80\}, and the mean squared error of the target image is approximately 5, which is an acceptable loss value with respect to \({\hat{I}}_i^L\).

%%% revision-start
\begin{figure*}[tbp]
  \centering
  \mbox{}
  \includegraphics[width=.9\linewidth]{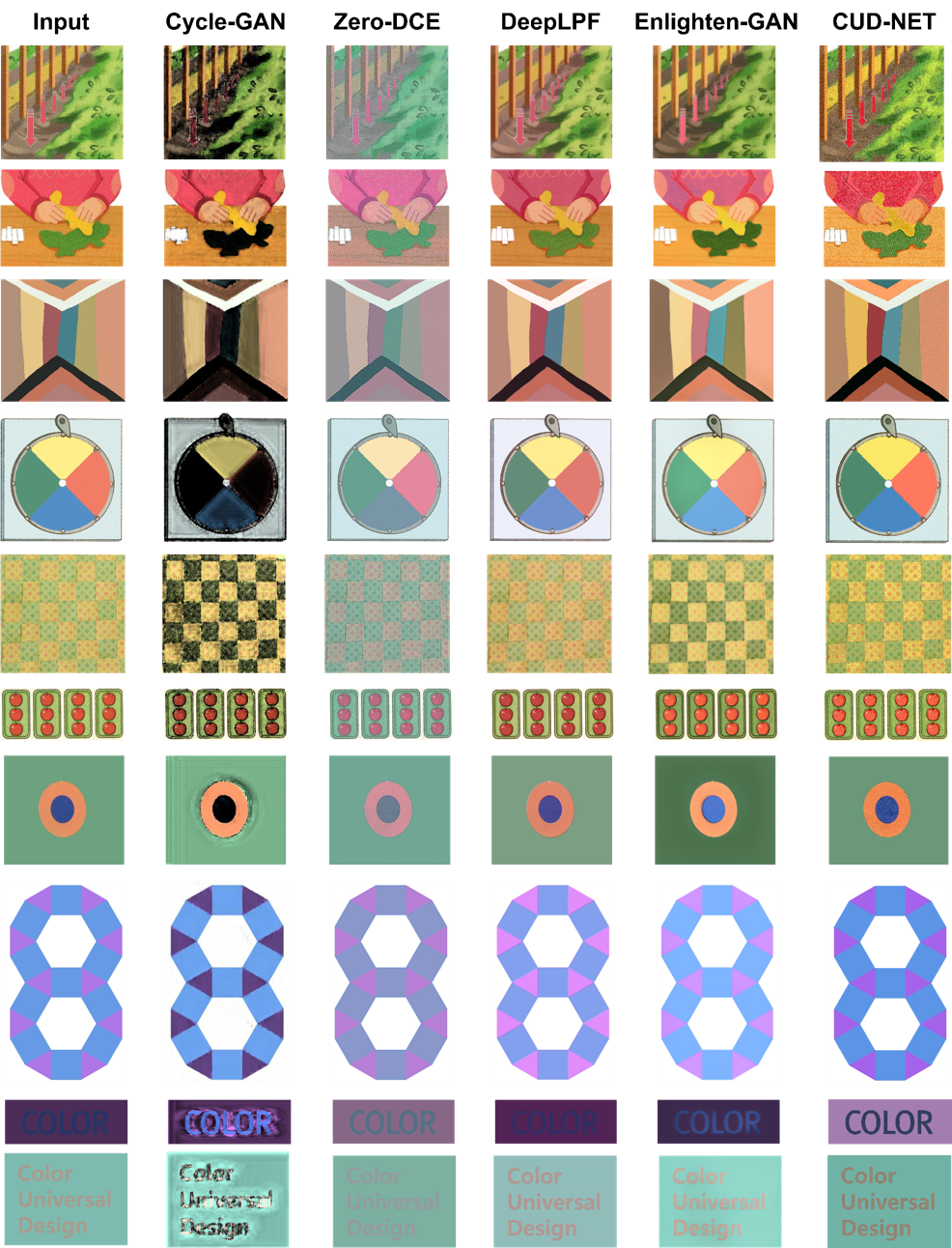}
  \caption{\label{fig:07-1}
           Comparison of predicted images in deuteranopia vision.}
\end{figure*}

\begin{figure*}[tbp]
  \centering
  \mbox{}
  \includegraphics[width=.9\linewidth]{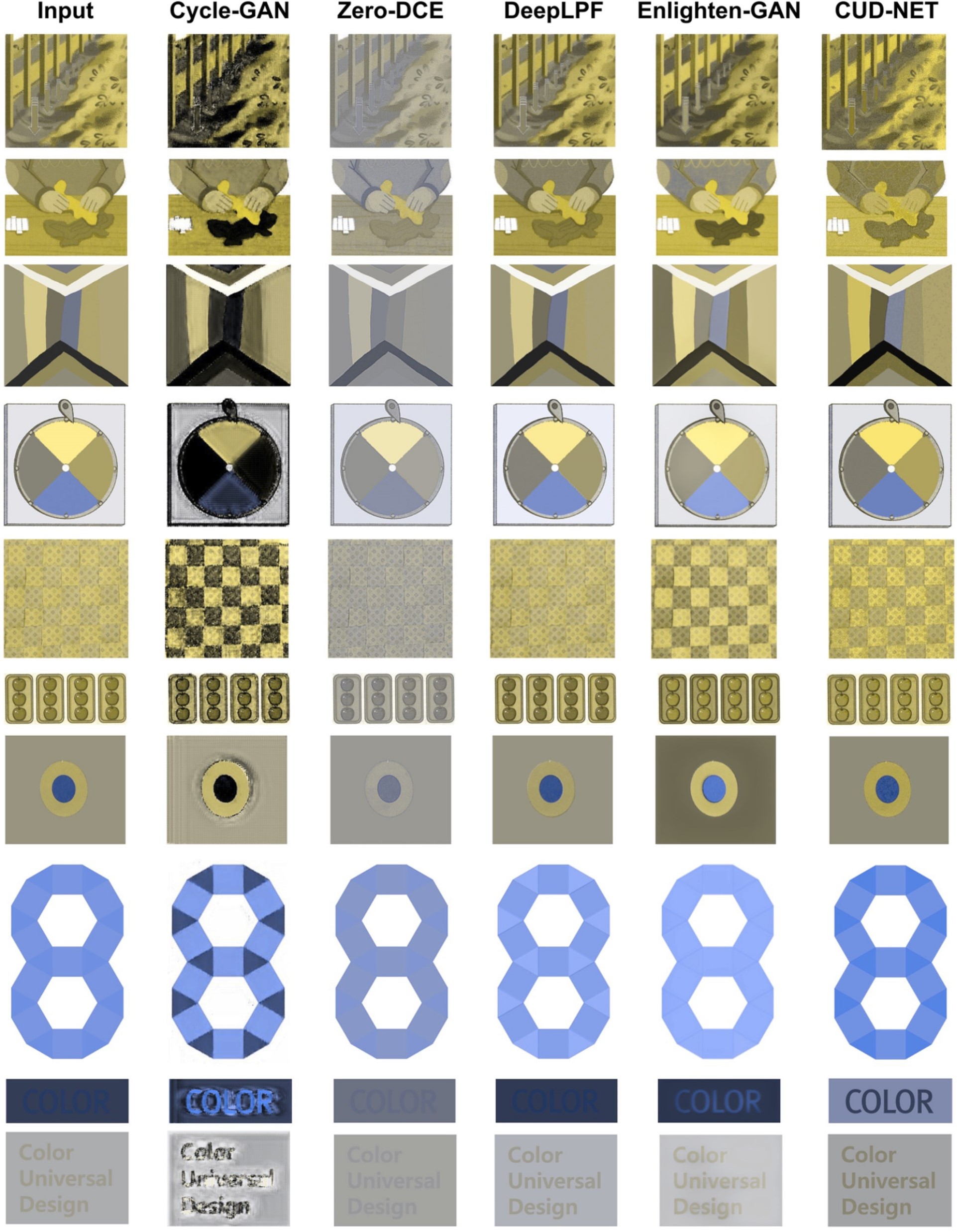}
  \caption{\label{fig:07-2}
           Comparison of predicted images in deuteranopia vision.}
\end{figure*}
%%% revision-end

\begin{table*}[tbp]
    \renewcommand{\arraystretch}{1.15}
    \centering
    \begin{tabular}{|c|c|c|c|c|c|c|}
    \hline 
    \textbf{Architecture} & \textbf{\(SSIM\left(\hat{I},\ I\right)\)} & \textbf{\(SSIM\left(\hat{I},\ T\right)\)} & \textbf{\(PSNR\left(\hat{I},\ I\right)\)} & \textbf{\(PSNR\left(\hat{I},\ T\right)\)} & \textbf{\(SSIM\mathrm{\cdot{}}MAE\)} & \textbf{\(PSNR\mathrm{\cdot{}}MAE\)} \\
    \hline
    \hline
    Cycle-GAN & 0.630 & 0.634 & 13.28 & 13.83 & 0.3191 & 8.4430 \\
    Zero-DCE & 0.924 & 0.888 & 21.95 & 18.67 & 0.0661 & 3.7300 \\
    DeepLPF & 0.850 & 0.831 & 26.31 & 20.34 & 0.1220 & 2.0566 \\
    Enlighten-GAN & 0.820 & 0.808 & 21.85 & 19.58 & 0.1470 & 3.9983 \\
    Enlighten-GAN (scaled) & \textbf{0.966} & 0.921 & 24.86 & \textbf{21.36} & 0.0392 & 3.4937 \\
    CUD-Net (low bottleneck feature) & 0.897 & 0.866 & 27.77 & 21.01 & 0.0901 & 2.0826 \\
    CUD-Net & 0.962 & \textbf{0.924} & \textbf{29.54} & 21.19 & \textbf{0.0312} & \textbf{1.4760} \\
    \hline
    \end{tabular}
  \caption{\label{tb:01}
       Evaluation results of comparison experiment. In the architecture column, Enlighten-GAN (scaled) indicates that the input image is downscaled to the predicted image. The CUD-Net with a low bottleneck feature achieved better results in the experiment of deuteranopia and protanopia subjects (Figure \ref{fig:09}); however, the evaluation metrics are lower than those of CUD-Net.}

\end{table*}

\textbf{Identity Loss}\cite{ZPIE17, TPW16}\textbf{.}
To capture the CUD objects in the model, we employ the identity loss \({L}_{identity}\left(T_i\right)\). In the case where the target image already satisfies the CUD, the application of the filter should be relatively weak compared to the input image. The input to the identity loss function is the target image, \(T_{ij}\), instead of the input image, \(I_{ij}\), and the reference of the loss function is also the target image, \(T_{ij}\), to maintain the value itself. While computing identity loss, we do not require conjugate prediction as we cannot judge the potential region using Equation 9.

\section{Experiments}

The hardware used in this experiment consisted of a Tesla V100 SXM2 and Intel Xeon Gold 5120, and the processing speed was approximately 40 images per minute. 
%%% revision-start
We used Adam optimizer with learning rate 1e-4, betas=\{0.9, 0.999\}, warm-up cosine scheduler with cycles every epochs, and batch size 2. 
%%% revision-end
To create the training dataset, Adobe Photoshop was employed to optimize the contrast in the L channel by adjusting the saturation and brightness of regions requiring color conversion based on deuteranopia vision simulation. Color experts refined approximately 1,500 vectorized images for the training data and 300 publication images for the 
validation data. 
%%% revision-start
As colors often appear distorted in images with rich gradations, the inference data used in publications primarily comprised vectorized images. We also carried out experiments using the MIT-Adobe FiveK dataset in deuteranopia vision, but the outcomes were heterogeneous, often resulting in color tone reversals leading to an all-black appearance. All the comparative experimental models in this study employed the same training, test, and validation data refined by color experts.

Figure \ref{fig:07-1} presents the results in descending order of the number of color combinations, with the top image having the more number of color combinations. The outcomes of CUD-Net exhibit a higher level of comprehensiveness compared to the input images within the context of deuteranopia vision. The CUD-Net not only discerns the areas of interest for non-CUD objects but also identifies CUD image where no adjustments are required, as exemplified in the 4th and 7th rows of Figure \ref{fig:07-2}.
%%% revision-end
Table 1 presents the evaluation of structure similarity (SSIM) \cite{ZBSS04} and peak signal-to-noise ratio (PSNR) metrics, which indicate the suitability of the image for CUD. It is important to note that the comparative models with lower metrics are more sensitive to high-gradation input images, which lead to color-heterogeneous images. Although SSIM and PSNR can measure the increase in contrast relative to the target image, they cannot assess whether the color preservation is adequately maintained. To address this, we evaluated SSIM and PSNR with three references: input images\(I\), predicted images \(\hat{I}\), and target images \(T\). The evaluation of  \(I\) and \(\hat{I}\) estimations reflects the effectiveness of the color preservation factor, while the evaluation of \(\hat{I}\) and \(T\)estimations represents the increase in contrast.

\[ SSIM\mathrm{\cdot}MAE=\ \frac{1}{N}\sum_{i=1}^{N}\left|\ SSIM\left({\hat{I}}_i,\ {T}_i\right)-SSIM\left({I}_i,\ {T}_i\right)\ \right| \ \ \ \ (10) \]

\[ PSNR\mathrm{\cdot}MAE=\ \frac{1}{N}\sum_{i=1}^{N}\left|\ PSNR\left({\hat{I}}_i,\ {T}_i\right)-PSNR\left({I}_i,\ {T}_i\right)\ \right| \ \ \ (11) \]

Additionally, we define the SSIM mean absolute error and PSNR mean absolute error to measure the extent of conversion between \(F: I\rightarrow T\) and \(F: I\rightarrow\ \hat{I}\) in Equations 10 and 11, similar with computing deviation. \(N\) is the total number of inference data points.

Among the compared models, Cycle-GAN and Zero-DCE delivered less satisfactory outcomes. Cycle-GAN's capacity to reconstruct object geometry was inadequate, and the image displayed low saturation and brightness, making the color shift nearly gray scale. Zero-DCE had a faded color and did not show a significant contrast difference from the input image. In general, both models failed to maintain color preservation, which was the main objective of our study.

\begin{figure}[htb]
  \centering
  \includegraphics[width=.9\linewidth]{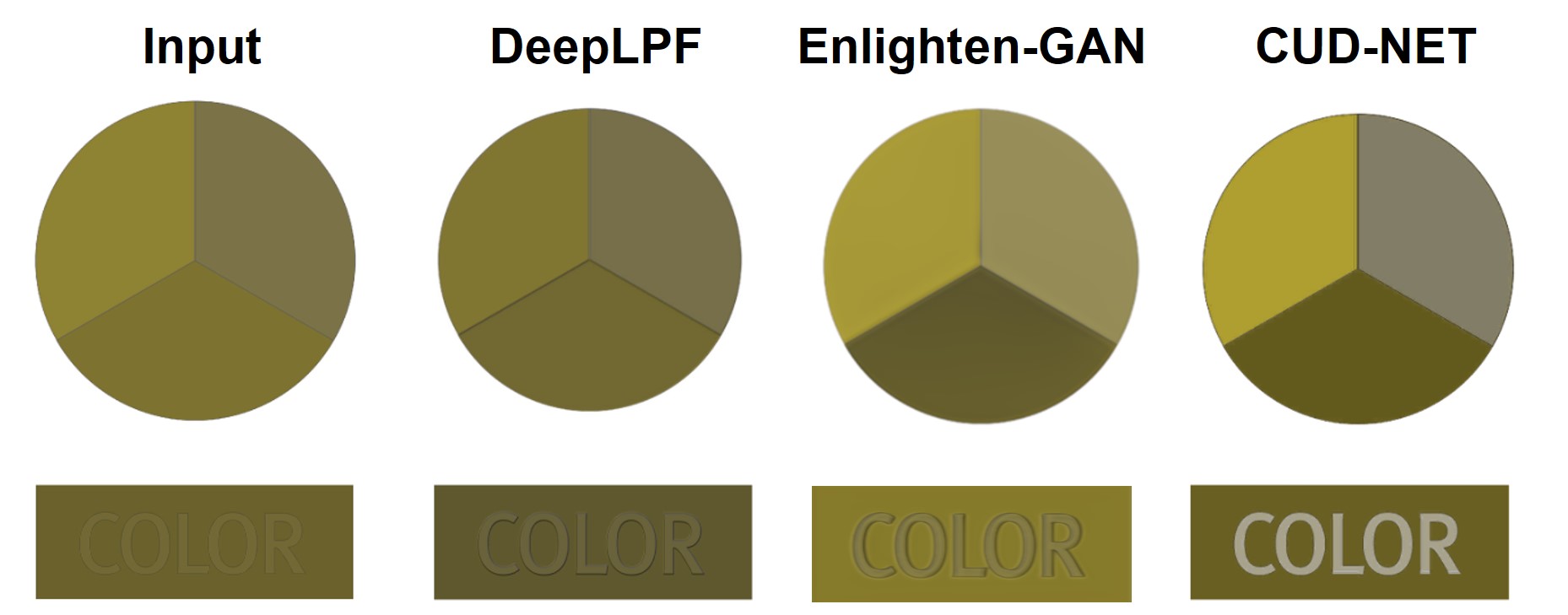}
  \caption{\label{fig:08}
           Qualitative evaluation of CUD suitability in deuteranopia vision.}
\end{figure}

On the other hand, DeepLPF successfully satisfied both color preservation and contrast requirements, as we aimed to achieve. Nonetheless, Figure \ref{fig:08} shows that DeepLPF tends to over-stabilize color filtering for images with few color combinations. Although color preservation was better than other experiments, some unsuccessful results from the perspective of contrast occurred due to DeepLPF's over-stable filtering.

Remarkably, the predicted images of Enlighten-GAN yielded reasonable results. However, simple color combinations or images that already satisfy the CUD often yielded results degenerated with low CUD suitability. Enlighten-GAN was able to generate results that we targeted; however, the deviation of the filter in it is so high that it sometimes fails to satisfy the contrast even on simple images or decreases the contrast. Owing to the problem of the GAN-based method including Enlighten-GAN, the model fixes the width and height of the predicted image. With the width and height of \(T\) and \(I\) downscaled to the size of Enlighten-GAN, \(\hat{I}\)(approximately 25k pixels in this experiment), the \(SSIM\left(\hat{I},\ I\right)\) and \(PSNR\left(\hat{I},\ T\right)\) yielded higher estimation in some metrics than CUD-Net. In the case opposite of \(\hat{I}\), up-scaled to a size of \(T\) and \(I\), a significantly low estimation was recorded because of the information loss of the up-scaling problem.

The results of Enlighten-GAN for predicted images were reasonable, but simple color combinations or images that already satisfied the CUD often produced degenerated results with low CUD suitability. While Enlighten-GAN targeted the generation of suitable results, the deviation of the filter was sometimes high, leading to failures in satisfying the contrast even on simple images or reducing the contrast. The limitations of the GAN-based method, including Enlighten-GAN, were that the model fixed the width and height of the predicted image. When the width and height of the input image \(I\) and target image \(T\) were downscaled to the size of Enlighten-GAN, approximately 25k pixels in this experiment, the \(SSIM\left(\hat{I},\ I\right)\) and \(PSNR\left(\hat{I},\ T\right)\) metrics yielded higher estimation than those of CUD-Net. Conversely, when \(\hat{I}\)was upscaled to the size of \(I\) and \(T\), a significantly low estimation was recorded due to the information loss from the upscaling problem.

Compared to other experiments, CUD-Net exhibited stability and robustness in generating predicted images with both color preservation and increased contrast. When comparing the values of the corresponding regions in \(I\) and \(\hat{I}\), the model scaled two L channel values in opposite directions in most cases, with one increasing while the other decreasing. However, reducing the number of bottleneck features of the model led to relatively high deviation of filter scales with increasing number of color combinations. Overall, CUD-Net achieved the highest estimation scores for the four evaluation metrics. Moreover, since our model uses a neural filter instead of a generative model, there is no loss of information due to the scaling of the predicted images.

 \begin{figure}[htb]
  \centering
  \includegraphics[width=\linewidth]{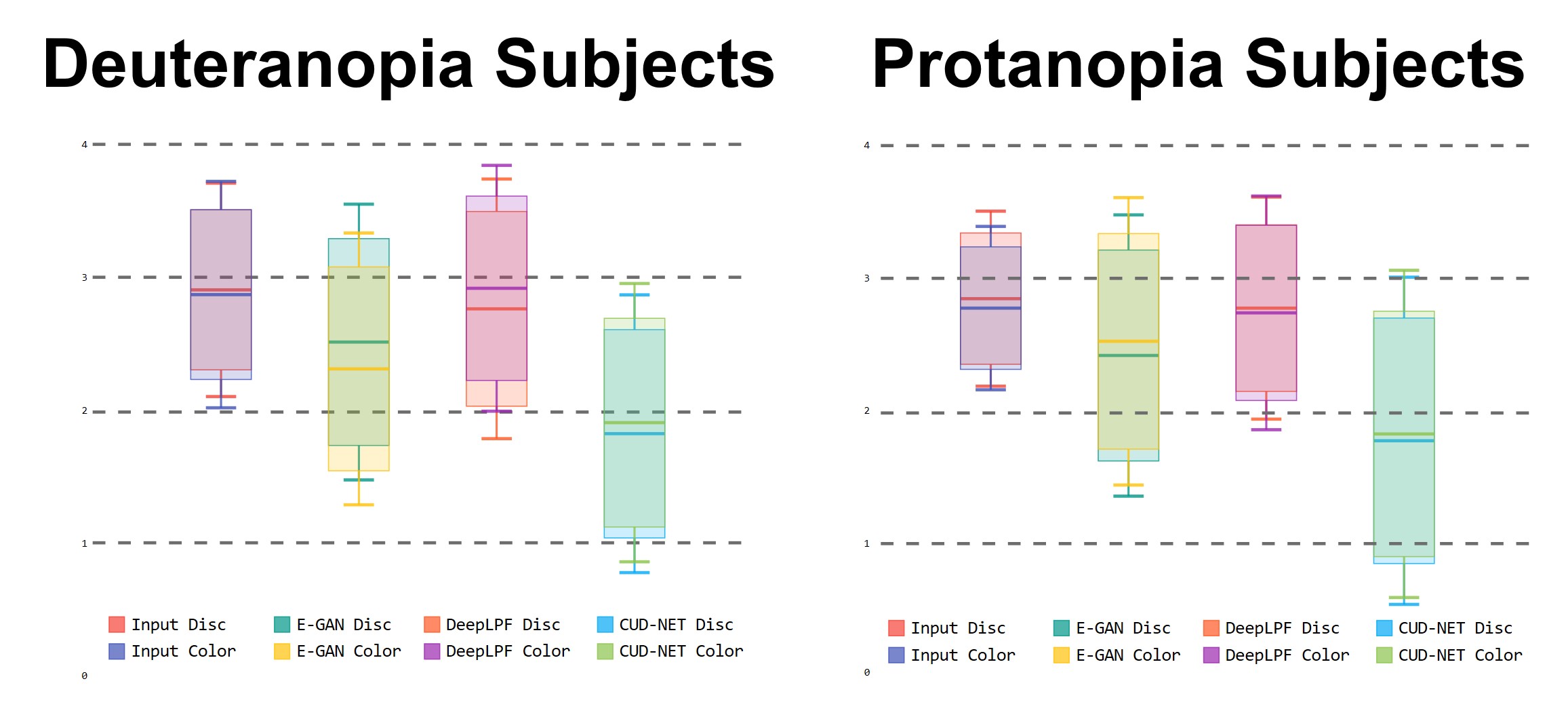}
  \caption{\label{fig:09}
           Evaluation of deuteranopia and protanopia. The box bar is ordered from left to right as input image \(I\), Enlighten-GAN, DeepLPF, and CUD-Net. The mean rank is represented by the y position of the box bar, and the deviation of each experiment is represented by the length of the box bar. Each experiment includes two box bars that indicate object distinguishability and color harmony. A lower position on the graph indicates a higher rank.}
\end{figure}

In Figure \ref{fig:09}, we present the evaluation of deuteranopia and protanopia by measuring object distinguishability and color harmony of the input image \(I\), the predicted image of Enlighten GAN, DeepLPF, and CUD-Net. To evaluate the performance, we conducted a user study involving six subjects, consisting of four deuteranomaly and two protanomaly subjects. They were asked to rank the four paired images based on their object distinguishability and color harmony, without knowing which model generated each image. The results showed that CUD-Net achieved the highest rank for object distinguishability among the deuteranopia subjects, with an average rank of 1.821, followed by Enlighten-GAN with an average rank of 2.512. Similarly, among the protanopia subjects, CUD-Net was ranked highest, followed by Enlighten-GAN, DeepLPF, and the input images. Furthermore, the evaluation of color harmony revealed that the subjects tended to prefer images with good object distinguishability. Among the six subjects, five preferred CUD-Net, while one subject had a subtle preference for Enlighten-GAN.

%%% revision-start
\section{Conclusion and future work}
%%% revision-end
The proposed neural network presented in this paper offers a solution for generating CUD images from non-CUD input images. The pre-processing step and multi-modality fusion layer aid in capturing the necessary information for color deficiency, while the conjugate loss function ensures greater adaptation to the CUD dataset. Compared to prior research, our method maintains high-resolution images and ensures stable color preservation and contrast through the use of neural filters per image. 
%%% revision-start
These results highlight the effectiveness of our approach and its potential to be applied in real-world scenarios where color vision deficiencies may impact visual perception on publications. In terms of practical applications, we are currently in the process of developing a web service and API package that leverages our CUD-Net for image conversion.
%%% revision-end

Our study has developed a reliable filter for single-colored vectorized images. However, it remains challenging to ensure stable and consistent results for real-world images that have a wide range of hues and shades. Our model's limitation is evident when certain pixel values are highly sensitive, leading to an increase in noise in the predicted image. 
%%% revision-start
Despite our experiments with the MIT-Adobe FiveK dataset in the context of deuteranopia vision, expanding the dataset from a different domain did not adequately encompass the diversity and complexity of real-world images for CUD applications.
%%% revision-end
To overcome this limitation, our future work aims to expand the dataset by introducing images with a more comprehensive range of hues and shades. Additionally, we will focus on improving the performance of the fusion layer to enhance the model's overall performance.
%%% revision-start
Our contribution is to broaden the scope of our research on color deficiencies to encompass aiding individuals with visual apprehensions. This research represents a cornerstone in advancing the field of color universal design through the pioneering application of machine learning methodologies.
%%% revision-end

% \section{Acknowledgement}
% This research is supported by Ministry of SMEs and Startups (Project Number: S3037924).

\bibliographystyle{IEEEtran}
\bibliography{bibliography.bib}

\begin{thebibliography}{10}
\providecommand{\url}[1]{#1}
\csname url@samestyle\endcsname
\providecommand{\newblock}{\relax}
\providecommand{\bibinfo}[2]{#2}
\providecommand{\BIBentrySTDinterwordspacing}{\spaceskip=0pt\relax}
\providecommand{\BIBentryALTinterwordstretchfactor}{4}
\providecommand{\BIBentryALTinterwordspacing}{\spaceskip=\fontdimen2\font plus
\BIBentryALTinterwordstretchfactor\fontdimen3\font minus
  \fontdimen4\font\relax}
\providecommand{\BIBforeignlanguage}[2]{{%
\expandafter\ifx\csname l@#1\endcsname\relax
\typeout{** WARNING: IEEEtran.bst: No hyphenation pattern has been}%
\typeout{** loaded for the language `#1'. Using the pattern for}%
\typeout{** the default language instead.}%
\else
\language=\csname l@#1\endcsname
\fi
#2}}
\providecommand{\BIBdecl}{\relax}
\BIBdecl

\bibitem{Won11}
B.~Wong, ``Points of view: Color blindnesss,'' pp. 409--426, 2011.

\bibitem{VZCR20}
\BIBentryALTinterwordspacing
V.~N. Varikuti, C.~Zhang, B.~Clair, and A.~L. Reynolds, ``Effect of enchroma
  glasses on color vision screening using ishihara and farnsworth d-15 color
  vision tests,'' \emph{Journal of American Association for Pediatric
  Ophthalmology and Strabismus}, vol.~24, no.~3, pp. 157.e1--157.e5, 2020.
  [Online]. Available:
  \url{https://www.sciencedirect.com/science/article/pii/S1091853120301002}
\BIBentrySTDinterwordspacing

\bibitem{ZhuMao2021}
Z.~Zhu and X.~Mao, ``Image recoloring for color vision deficiency compensation:
  a survey,'' \emph{The Visual Computer}, vol.~37, pp. 1--20, 12 2021.

\bibitem{WEG87}
\BIBentryALTinterwordspacing
S.~Wold, K.~Esbensen, and P.~Geladi, ``Principal component analysis,''
  \emph{Chemometrics and Intelligent Laboratory Systems}, vol.~2, no.~1, pp.
  37--52, 1987, proceedings of the Multivariate Statistical Workshop for
  Geologists and Geochemists. [Online]. Available:
  \url{https://www.sciencedirect.com/science/article/pii/0169743987800849}
\BIBentrySTDinterwordspacing

\bibitem{TSC20}
A.~Tao, K.~Sapra, and B.~Catanzaro, ``Hierarchical multi-scale attention for
  semantic segmentation,'' 2020.

\bibitem{ZGL*20}
B.~Zoph, G.~Ghiasi, T.-Y. Lin, Y.~Cui, H.~Liu, E.~D. Cubuk, and Q.~V. Le,
  ``Rethinking pre-training and self-training,'' 2020.

\bibitem{kirillov2023segment}
A.~Kirillov, E.~Mintun, N.~Ravi, H.~Mao, C.~Rolland, L.~Gustafson, T.~Xiao,
  S.~Whitehead, A.~C. Berg, W.-Y. Lo, P.~Dollár, and R.~Girshick, ``Segment
  anything,'' 2023.

\bibitem{AHB*18}
P.~Anderson, X.~He, C.~Buehler, D.~Teney, M.~Johnson, S.~Gould, and L.~Zhang,
  ``Bottom-up and top-down attention for image captioning and visual question
  answering,'' in \emph{Proceedings of the IEEE Conference on Computer Vision
  and Pattern Recognition (CVPR)}, June 2018.

\bibitem{KZG*17}
\BIBentryALTinterwordspacing
R.~Krishna, Y.~Zhu, O.~Groth, J.~Johnson, K.~Hata, J.~Kravitz, S.~Chen,
  Y.~Kalantidis, L.-J. Li, D.~A. Shamma, M.~S. Bernstein, and L.~Fei-Fei,
  ``Visual genome: Connecting language and vision using crowdsourced dense
  image annotations,'' \emph{Int. J. Comput. Vision}, vol. 123, no.~1, p.
  32–73, May 2017. [Online]. Available:
  \url{https://doi.org/10.1007/s11263-016-0981-7}
\BIBentrySTDinterwordspacing

\bibitem{LYL*20}
X.~Li, X.~Yin, C.~Li, P.~Zhang, X.~Hu, L.~Zhang, L.~Wang, H.~Hu, L.~Dong,
  F.~Wei, Y.~Choi, and J.~Gao, ``Oscar: Object-semantics aligned pre-training
  for vision-language tasks,'' in \emph{Computer Vision -- ECCV 2020},
  A.~Vedaldi, H.~Bischof, T.~Brox, and J.-M. Frahm, Eds.\hskip 1em plus 0.5em
  minus 0.4em\relax Cham: Springer International Publishing, 2020, pp.
  121--137.

\bibitem{RG19a}
M.~G. Ribeiro and A.~Gomes, ``Recoloring algorithms for colorblind people,''
  \emph{ACM Computing Surveys (CSUR)}, vol.~52, pp. 1 -- 37, 2019.

\bibitem{IMPS06}
\BIBentryALTinterwordspacing
G.~Iaccarino, D.~Malandrino, M.~Del~Percio, and V.~Scarano, ``Efficient
  edge-services for colorblind users,'' in \emph{Proceedings of the 15th
  International Conference on World Wide Web}, ser. WWW '06.\hskip 1em plus
  0.5em minus 0.4em\relax New York, NY, USA: Association for Computing
  Machinery, 2006, p. 919–920. [Online]. Available:
  \url{https://doi.org/10.1145/1135777.1135944}
\BIBentrySTDinterwordspacing

\bibitem{FRGG13}
\BIBentryALTinterwordspacing
D.~R. Flatla, K.~Reinecke, C.~Gutwin, and K.~Z. Gajos, \emph{SPRWeb: Preserving
  Subjective Responses to Website Colour Schemes through Automatic
  Recolouring}.\hskip 1em plus 0.5em minus 0.4em\relax New York, NY, USA:
  Association for Computing Machinery, 2013, p. 2069–2078. [Online].
  Available: \url{https://doi.org/10.1145/2470654.2481283}
\BIBentrySTDinterwordspacing

\bibitem{KWK21}
\BIBentryALTinterwordspacing
M.~Kumar, D.~Weissenborn, and N.~Kalchbrenner, ``Colorization transformer,'' in
  \emph{International Conference on Learning Representations}, 2021. [Online].
  Available: \url{https://openreview.net/forum?id=5NA1PinlGFu}
\BIBentrySTDinterwordspacing

\bibitem{IZZE17}
P.~Isola, J.-Y. Zhu, T.~Zhou, and A.~A. Efros, ``Image-to-image translation
  with conditional adversarial networks,'' in \emph{Proceedings of the IEEE
  Conference on Computer Vision and Pattern Recognition (CVPR)}, July 2017.

\bibitem{PEZZ20}
T.~Park, A.~A. Efros, R.~Zhang, and J.-Y. Zhu, ``Contrastive learning for
  unpaired image-to-image translation,'' in \emph{Computer Vision -- ECCV
  2020}, A.~Vedaldi, H.~Bischof, T.~Brox, and J.-M. Frahm, Eds.\hskip 1em plus
  0.5em minus 0.4em\relax Cham: Springer International Publishing, 2020, pp.
  319--345.

\bibitem{JGL*21}
Y.~{Jiang}, X.~{Gong}, D.~{Liu}, Y.~{Cheng}, C.~{Fang}, X.~{Shen}, J.~{Yang},
  P.~{Zhou}, and Z.~{Wang}, ``Enlightengan: Deep light enhancement without
  paired supervision,'' \emph{IEEE Transactions on Image Processing}, vol.~30,
  pp. 2340--2349, 2021.

\bibitem{WZF*19}
R.~Wang, Q.~Zhang, C.-W. Fu, X.~Shen, W.-S. Zheng, and J.~Jia, ``Underexposed
  photo enhancement using deep illumination estimation,'' in \emph{Proceedings
  of the IEEE/CVF Conference on Computer Vision and Pattern Recognition
  (CVPR)}, June 2019.

\bibitem{DLT18}
\BIBentryALTinterwordspacing
Y.~Deng, C.~C. Loy, and X.~Tang, ``Aesthetic-driven image enhancement by
  adversarial learning,'' in \emph{Proceedings of the 26th ACM International
  Conference on Multimedia}, ser. MM '18.\hskip 1em plus 0.5em minus
  0.4em\relax New York, NY, USA: Association for Computing Machinery, 2018, p.
  870–878. [Online]. Available: \url{https://doi.org/10.1145/3240508.3240531}
\BIBentrySTDinterwordspacing

\bibitem{BCPS19}
S.~Bianco, C.~Cusano, F.~Piccoli, and R.~Schettini, ``Content-preserving tone
  adjustment for image enhancement,'' in \emph{Proceedings of the IEEE/CVF
  Conference on Computer Vision and Pattern Recognition (CVPR) Workshops}, June
  2019.

\bibitem{GLG*20}
C.~Guo, C.~Li, J.~Guo, C.~C. Loy, J.~Hou, S.~Kwong, and R.~Cong,
  ``Zero-reference deep curve estimation for low-light image enhancement,'' in
  \emph{Proceedings of the IEEE/CVF Conference on Computer Vision and Pattern
  Recognition (CVPR)}, June 2020.

\bibitem{MMM*20}
S.~Moran, P.~Marza, S.~McDonagh, S.~Parisot, and G.~Slabaugh, ``Deeplpf: Deep
  local parametric filters for image enhancement,'' in \emph{Proceedings of the
  IEEE/CVF Conference on Computer Vision and Pattern Recognition (CVPR)}, June
  2020.

\bibitem{RG19b}
\BIBentryALTinterwordspacing
M.~Ribeiro and A.~J.~P. Gomes, ``Recoloring algorithms for colorblind people: A
  survey,'' \emph{ACM Comput. Surv.}, vol.~52, no.~4, Aug. 2019. [Online].
  Available: \url{https://doi.org/10.1145/3329118}
\BIBentrySTDinterwordspacing

\bibitem{HMKO19}
\BIBentryALTinterwordspacing
S.~Hira, A.~Matsumoto, K.~Kihara, and S.~Ohtsuka, ``Hue rotation (hr) and hue
  blending (hb): Real-time image enhancement methods for digital component
  video signals to support red-green color-defective observers,'' \emph{Journal
  of the Society for Information Display}, vol.~27, no.~7, pp. 409--426, 2019.
  [Online]. Available:
  \url{https://onlinelibrary.wiley.com/doi/abs/10.1002/jsid.758}
\BIBentrySTDinterwordspacing

\bibitem{DTAA09}
P.~Doliotis, G.~Tsekouras, C.-N. Anagnostopoulos, and V.~Athitsos,
  ``Intelligent modification of colors in digitized paintings for enhancing the
  visual perception of color-blind viewers,'' in \emph{Artificial Intelligence
  Applications and Innovations III}, Iliadis, Maglogiann, Tsoumakasis,
  Vlahavas, and Bramer, Eds.\hskip 1em plus 0.5em minus 0.4em\relax Boston, MA:
  Springer US, 2009, pp. 293--301.

\bibitem{LCY14}
M.~Lin, Q.~Chen, and S.~Yan, ``Network in network,'' 2014.

\bibitem{FPY*16}
A.~Fukui, D.~H. Park, D.~Yang, A.~Rohrbach, T.~Darrell, and M.~Rohrbach,
  ``Multimodal compact bilinear pooling for visual question answering and
  visual grounding,'' 2016.

\bibitem{CHARIKAR20043}
\BIBentryALTinterwordspacing
M.~Charikar, K.~Chen, and M.~Farach-Colton, ``Finding frequent items in data
  streams,'' \emph{Theoretical Computer Science}, vol. 312, no.~1, pp. 3--15,
  2004, automata, Languages and Programming. [Online]. Available:
  \url{https://www.sciencedirect.com/science/article/pii/S0304397503004006}
\BIBentrySTDinterwordspacing

\bibitem{MMS19}
S.~Moran, S.~McDonagh, and G.~Slabaugh, ``Curl: Neural curve layers for global
  image enhancement,'' 2019.

\bibitem{WSB03}
Z.~{Wang}, E.~P. {Simoncelli}, and A.~C. {Bovik}, ``Multiscale structural
  similarity for image quality assessment,'' in \emph{The Thrity-Seventh
  Asilomar Conference on Signals, Systems Computers, 2003}, vol.~2, 2003, pp.
  1398--1402 Vol.2.

\bibitem{SAC*17}
K.~T. Schütt, F.~Arbabzadah, S.~Chmiela, K.~R. Müller, and A.~Tkatchenko,
  ``Quantum-chemical insights from deep tensor neural networks,'' \emph{Nature
  Communications}, vol.~8, no.~1, 2017.

\bibitem{ZPIE17}
J.-Y. Zhu, T.~Park, P.~Isola, and A.~A. Efros, ``Unpaired image-to-image
  translation using cycle-consistent adversarial networks,'' in
  \emph{Proceedings of the IEEE International Conference on Computer Vision
  (ICCV)}, Oct 2017.

\bibitem{TPW16}
Y.~Taigman, A.~Polyak, and L.~Wolf, ``Unsupervised cross-domain image
  generation,'' 2016.

\bibitem{ZBSS04}
{Zhou Wang}, A.~C. {Bovik}, H.~R. {Sheikh}, and E.~P. {Simoncelli}, ``Image
  quality assessment: from error visibility to structural similarity,''
  \emph{IEEE Transactions on Image Processing}, vol.~13, no.~4, pp. 600--612,
  2004.

\end{thebibliography}

\end{document}